\documentclass[10pt,preprint2]{aastex}
\usepackage{xspace}
\usepackage{amsmath}
\def\figdir{/idunn/scratch2/gnedin/PAPERS/FIGS/LGD}
\def\figdir{.}

\def\ie{{\it i.e.}}
\def\eg{{\it e.g.}}
\newcommand{\ltsima}{\mbox{$\; \buildrel < \over \sim \;$}}
\def\simlt{\lower.5ex\hbox{\ltsima}}
\def\gtsima{$\; \buildrel > \over \sim \;$}
\def\simgt{\lower.5ex\hbox{\gtsima}}

\def\HI{\mbox{\ion{H}{1}}}
\def\HII{\mbox{\ion{H}{2}}}
\def\GI{\mbox{\ion{He}{1}}}
\def\GII{\mbox{\ion{He}{2}}}

\def\H2{\mbox{H$_2$}\xspace}

\def\dim#1{\mbox{\,#1}}
\def\figname#1{\figdir/#1}

\def\hide#1{}

\def\nameIIplu{``survivors''}

\def\new#1{{#1}}

\begin{document}


\title{Formation Histories of Dwarf Galaxies in the Local Group}

\author{Massimo Ricotti\altaffilmark{1} and Nickolay Y.\
  Gnedin\altaffilmark{2}}  
\altaffiltext{1}{Institute of Astronomy, Madingley Road, Cambridge CM3 0HA,
  England; ricotti@ast.cam.ac.uk}
\altaffiltext{2}{CASA, University of Colorado, Boulder, CO 80309, USA;
gnedin@casa.colorado.edu}

\begin{abstract}
We compare the properties of dwarf galaxies in the Local Group with
  the simulated galaxies formed before reionization in a cosmological
  simulation of unprecedented spatial and mass resolution, including
  radiative feedback effects. We find that a subset of the Local Group
  dwarfs are remarkably similar to the simulated dwarf galaxies in all
  their properties already before reionization.
  Based on this similarity, we propose the hypothesis that Local Group
  dwarfs form in a variety of ways: some of them are ``true fossils''
  of the pre-reionization era, some of them form most of their stars
  later, after reionization (we call them \nameIIplu\ of the
  reionization era), and the rest of them form an intermediate group
  of ``polluted fossils''. We also identify a simple observational
  test that is able to falsify our hypothesis.
\end{abstract}


\section{Introduction}

Nothing emphasizes our lack of understanding of the origin of the
dwarf galaxies of the Local Group more than the well known
``substructure crisis'' of the Cold Dark Matter (CDM) paradigm (Klypin
et al.\ 1999; Moore et al.\ 1999): the fact that the number of
dark matter halos in the Local Group predicted by CDM simulations is
much larger than the number of observed Galactic satellites. Several
explanations have been suggested so far for resolving the ``crisis''.

The hypothesis of Bullock, Kravtsov, \& Weinberg (2000) - we will call
it hereafter the ``reionization scenario'' - proposed that dwarf
spheroidal (dSph) galaxies in the Local Group formed during the
pre-reionization era and processes of photo-ionization feedback
(Babul \& Rees 1992; Efstathiou 1992; Shapiro, Giroux, \& Babul 1994;
Haiman, Rees, \& Loeb 1996; Thoul \& Weinberg 1996; Quinn, Katz, \&
Efstathiou 1996; Weinberg, Hernquist, \& Katz 1997; Navarro \&
Steinmetz 1997; Gnedin 2000b; Dijkstra et al.\ 2004; Shapiro, Iliev,
\& Raga 2004, Susa \& Umemura 2004) resulted in the suppression of
star formation below the observable level in 90\% of the low mass dark
matter halos populating the Local Group. The ``reionization scenario''
is prompted by the fact that almost all Local Group dwarf Spheroidals
exhibit a prominent old population and a decline of their star
formation rates about $10\dim{Gyr}$ ago. While in the standard
$\Lambda$-CDM theory (Spergel et al.\ 2003, Tegmark et al.\ 2004)
stellar reionization took place $12.5\dim{Gyr}$ ago, current
observational techniques for measuring absolute stellar ages are not
able to differentiate between $10$ and $12.5\dim{Gyr}$ (\eg, Krauss \&
Chaboyer\ 2003). Although the observed drop of star formation rate is
consistent with reionization at redshift $z \sim 6$, as noted by
Grebel \& Gallagher\ 2004, the metallicity spreads observed in the old
populations may imply a star formation history protracted over about 
$2\dim{Gyr}$, to redshift $z \sim 3$. In this paper we
critically discuss this important issue. We will show that our
simulations reproduce the observed metallicity spreads without a
protracted period of star formation. The metallicity spreads observed
in our simulated dwarfs can be understood in terms of hierarchical
accretion of sub-halos containing stars with different metallicities.

An alternative explanation to the ``substructure crisis'' was proposed
by Stoehr et al.\ (2002). According to their hypothesis, dwarf
galaxies are hosted in more massive dark matter halos ($\sim
10^9-10^{10}$ M$_\odot$) than in the reionization scenario. While this
idea resolves the ``crisis'', it does not provide an explanation for
the origin of dwarf galaxies. 

A more comprehensive and complex picture has been recently proposed by 
Kravtsov, Gnedin, \& Klypin (2004).
Under their hypothesis - which we will hereafter call the ``tidal
scenario'' - most dwarf galaxies used to be much (up to a factor of
100) more massive during the formation episodes of the Milky Way and
Andromeda, and so were not affected by photo-ionization feedback. They
formed most of their stars at $z\sim 3$ but later most of the dark
matter was tidally stripped during their evolution within the Milky
Way and Andromeda halos. Kravtsov et al.\ (2004) emphasize that in
their model a large variety of star formation histories is observed,
and some of dwarf galaxies do indeed follow the ``reionization
track'', but the bulk of stars in the Local Group dwarfs are formed after
reionization. 

Thus, the key difference between the reionization and tidal hypotheses is
whether most of the stars in a given dwarf galaxy formed in {\em halos
  of mass $\simlt 10^8$ M$_\odot$} before reionization ($z>6$,
$12.5-13\dim{Gyr}$ ago) or in {\em more massive halos} after
reionization, during the epoch of formation of the Milky Way and
Andromeda ($z\sim 2-3$, $10-11\dim{Gyr}$ ago).

In this paper we propose that dSphs are fossils of galaxies with mass
$M_{\rm dm}<10^8$ M$_\odot$ that formed before reionization while dIrr
formed later in more massive halos.  The original contribution of our
study is that for the first time we have been able to simulate the
first population of galaxies including the relevant feedback processes
that are known to self-limit their formation. Before our work
(Ricotti, Gnedin, \& Shull 2002a, 2002b) 
the mainstream wisdom was that negative feedback
effects (namely the rapid photo-dissociation of \H2) suppress the
formation of all galaxies with mass smaller than $10^8$ solar masses.
We find instead that these galaxies do form and, at the epoch of their
formation, they closely resemble dSph galaxies. This second result is
the one we emphasize in the present paper. We also show that
reionization feedback is not the dominant mechanism that suppresses
star formation in most small mass halos. Negative feedbacks such as
photo-heating by the stars inside each galaxy and inefficient cooling
produce the observed properties of dSphs well before reionization of
the IGM is complete.

It is quite surprising that we find such an astonishing agreement
between all properties of simulated dwarf galaxies at $z \sim 8$ and
dSphs. Our simulations did improve on previous ones but are still far
from a complete treatment of the problem that, as we will briefly
illustrate, is non-trivial. The efficiency of star formation in the
first mini-halos with masses $\simlt 10^8$ M$_\odot$ (\ie, virial
temperature $T_{\rm vir} \simlt 10^4$ K) is determined by many
competing feedback processes, including the formation and disruption
of \H2, metal enrichment, photoevaporation and SN explosions.  Since
small mass proto-galaxies cannot cool by Lyman-alpha emission, the
presence of \H2 molecules is essential to trigger their initial
collapse.  We know that \H2 is easily destroyed by FUV radiation
emitted by the first stars (negative feedback). On the other hand
ionizing radiation and dust production are known to enhance the
formation rate of \H2 (positive feedbacks).  Whether or not efficient
star formation is possible in these mini-halos depends on which
feedback (positive or negative) is dominant.  By including 3D
radiative transfer of ionizing radiation (Ricotti et al 2002a) in the
simulations we are able to model two additional feedbacks: the
positive feedback for \H2 formation and the photoevaporation of dwarfs
from internal sources.  As already mentioned, contrary to previous
studies we find that low surface brightness galaxies do form in a
fraction of the first mini-halos.  Reionization does not affect
substantially their properties because they have already lost most of
their gas due to internal ionizing sources.  After reionization we
expect that the formation of small mass galaxies will stop and their
stellar populations will evolve only passively.  Only the few most
massive dwarfs in our simulation retained some of their gas and may
continue to form stars after reionization.

In summary the goal of this paper is to provide evidences in support
of the primordial origin of dSphs by comparing the abundant
observational data on Local Group dwarfs with the results of detailed
numerical simulation of formation of dwarf galaxies in the
pre-reionization era.  Aside for the origin of dSph galaxies the aim
of study is to improve our understanding on theories of the formation
of the first galaxies and answer a longstanding question: how massive
were the first galaxies?

\section{Simulation}

The results shown in the present work are based on a detailed analysis
of the galaxies found in high-resolution cosmological simulations of a
cosmological volume during the pre-ionization era.  The simulations
include dark matter, gas, stars and the feedback of star formation on
the gas chemistry and cooling. The main addition with respect to other
works (\eg, Machacek et al. 2001, Tassis et al. 2003) is a
self-consistent treatment of the spatially-inhomogeneous and time
dependent radiative transfer of photons in four frequency bands (\HI
ionizing radiation, \GI\ and \GII\ ionizing radiation and optically thin
radiation). This allows us to study in addition to the global feedback
effects of the background radiation, also the role of internal and
local feedback effects.  Partially because of the inclusion of a more
realistic treatment of radiative feedback, we find that the \H2
photodissociating background is not the dominant feedback and does not
stop star formation in mini-halos.

The simulation we use in this paper has been thoroughly described in
Ricotti et al.\ (2002a, 2002b) as run ``256L1p3''. Here we
just remind the reader that the simulation included $256^3$ dark
matter particles, an equal number of baryonic cells, and more than
700,000 stellar particles. The mass resolution of our simulation is
900 solar masses in baryons, and real comoving spatial resolution
(twice the Plummer softening length) is $150h^{-1}\dim{pc}$ (which
corresponds to a physical scale of 24 parsecs at $z=8.3$). This
unprecedented resolution allows us to resolve the cores of all
simulated galaxies that would correspond to the observed Local Group
dwarfs.  The simulation includes most of the relevant physics,
including time-dependent spatially-variable radiative transfer,
detailed radiative transfer in Lyman-Werner bands, non-equilibrium
ionization balance, etc.

In this work, however, we also include the effect of reionization in
that simulation. Because the size of the simulation box has been fixed
at $1h^{-1}$ comoving Mpc, the simulation volume is too small to model
the process of cosmological reionization with sufficient accuracy. We,
therefore, assume that the simulation volume is located inside an
\HII\ region of a bright galaxy at a higher redshift. Specifically, we
introduce a source of ionizing radiation within the computational box,
properly biased, which corresponds to a star-forming galaxy with the
constant star formation rate of 1 solar mass per year (similar to star
formation rates of observed Lyman Break Galaxies at $z\sim4$, Steidel
et al.\ 1999). The source is switched on at $z=9.0$, and by $z=8.3$
the whole simulation box is completely ionized, and star formation in
dwarf galaxies has ceased.  Because of the computational expense, we
have not continued our simulation beyond $z=8.3$.  As mentioned in the
introduction the photoevaporation of dwarf galaxies after reionization
has been studied extensively. Recent works include those of Shapiro,
Iliev, \& Raga (2004) and Susa \& Umemura (2004) that addressed the
importance of self-shielding using 3D simulations coupled to 1D
radiative transfer (\ie, they assume that the ionizing source is very
far from the dwarf galaxy).

\begin{figure}[t]
\plotone{\figname{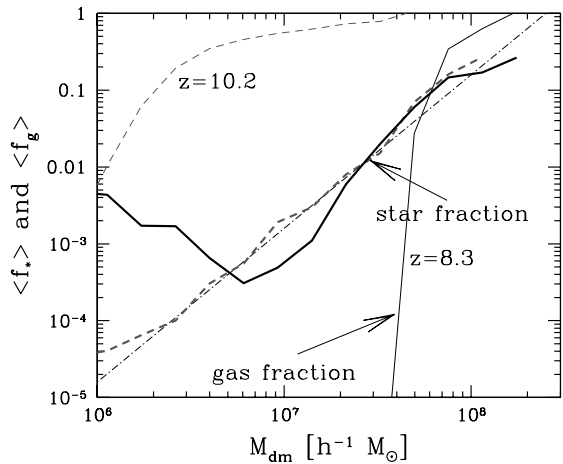}}
\caption{The mean gas fraction (thin lines) and the mean stellar fraction
  (thick lines) as a function of the dark matter mass for the simulated 
  objects at $z=10.2$ (dashed gray lines) and at the final moment of our
  simulation, $z=8.3$ (solid black lines). Both fractions are normalized by
  the universal baryon fraction.} 
\label{figMM}
\end{figure}
Figure \ref{figMM} shows the mean gas (thin lines) and stellar (thick
lines) fractions as a function of the dark matter mass for the objects
in our simulation. We show these quantities at redshift $z=10.2$
(dashed lines) and $z=8.3$ (solid lines).  Notice that the objects
with masses below about $10^8\dim{M}_\odot$ contain no gas at the end
of our simulation. Most of the small mass galaxies loose all their gas
well before reionization that in our simulation happens between
redshift $z \sim 9$ and $8.3$. This is the result of gas photo-heating
by massive stars inside the galaxy and not by the external ionizing
background. A remarkable result is that the mean stellar fraction is a
time-independent quantity.  It is apparent from the figure that the
mean stellar fraction as a function of halo mass is well approximated
by power law that does not evolve with redshift (the increase in the
stellar fraction for the lowest mass objects is probably an artifact
of the limited mass resolution).

\hide{ Only after introducing by hand a source of ionizing photons,
  before reionization is completed, we observe an increase in the mean
  stellar fraction for the lowest mass objects. There are a few
  explanations for this: it may be simply an artifact of the limited
  mass resolution (the smaller mass halos are resolved with 100-1000
  dark matter particles) or a real effect produced by an increase of
  their star formation combined with tidal mass loss of the outer part
  of their dark matter halos, similar to the ``tidal scenario''
  proposed by Kravtsov et al. (2004). In support of this second
  hypothesis we also observe, before reionization, a temporary
  increase of the formation rate of molecular hydrogen that could
  boost the star formation in some object. This ``positive feedback''
  process is probably the result of the increasing flux of the
  ionizing background before overlap of the \HII\ regions (Haiman,
  Rees, \& Loeb 1996; Ricotti, Gnedin \& Shull\ 2001).  }

\begin{figure}[t]
\plotone{\figname{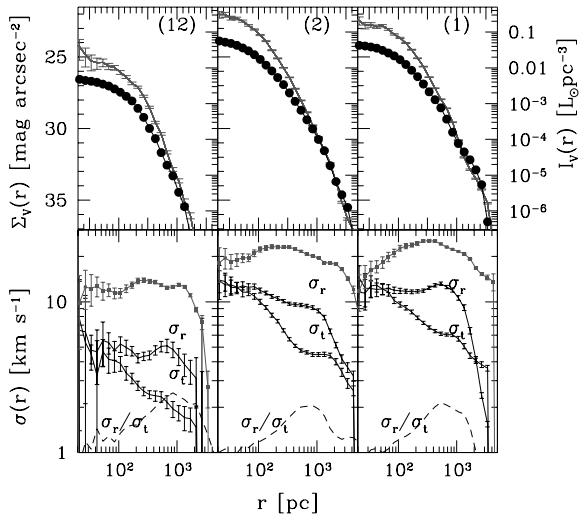}}
\caption{The stellar surface (filled circles) and volume (gray lines)
  densities (top row) and stellar (black lines) and dark matter (gray
  triangles) velocity dispersions (bottom row) as a function of radius
  for three representative objects from our simulation at $z=8.3$. The
  degree of anisotropy of the velocity dispersion is usually measured
  by the parameter $\beta=1-(\sigma_{r}/\sigma_{t})^{-2}$, where
  $\sigma_{r}$ is the radial velocity dispersion and $\sigma_{t}$ is
  tangential velocity dispersions. The dashed lines in the bottom
  panels show $\sigma_{r}/\sigma_{t}$.}
\label{figMP}
\end{figure}
Figure \ref{figMP} shows radial profiles for three representative and
well resolved objects. In the top panels we show surface (filled
circles) and volume (gray lines) luminosity densities in the V band.
The bottom panels show the velocity dispersions for both the dark
matter (about $50,000$ particles per halo) and stars (about $10,000$
particles per halo).  Notice that stellar velocity dispersions (black
points) are below that of the dark matter (gray triangles), and that
stellar orbits are mildly anisotropic in the outer parts. The dashed
lines in the bottom panels show $\sigma_{r}/\sigma_{t}$, where
$\sigma_{r}$ and $\sigma_{t}$ are the radial and tangential velocity
dispersions, respectively.

\hide{
The stellar velocity dispersion profile shown in figure \ref{figMP}
becomes increasingly unreliable in the inner parts due to the small
number of particles.  But in a previous study, Ricotti\ (2003) have
found indication of a dependence of the the inner density profile of
dark matter halos on their mass that could help to reconcile
observations with simulations: the halos of dwarf galaxies have a
shallower inner density profile than in normal galaxies and clusters.
Later, Ricotti \& Wilkinson\ (2004) have shown that dwarf halos
extracted from the simulation at $z=10$ and evolved in isolation are
dynamically stable and will maintain a core at redshift $z=0$. In this
case it was shown that the dark matter profiles would be consistent
with measures of the stellar velocity dispersion in the inner parts of
Ursa Minor and Draco dwarf Spheroidals. Recently, new data (Wilkinson
et al.\ 2004) show that the stellar velocity dispersion in the outer
parts of Ursa Minor decreases sharply. This measurement is difficult
to understand in models with isotropic velocity dispersion for the
stars, but the problem will certainly be alleviated assuming a mildly
tangential anisotropy in the outer parts as found in our
numerical simulations.
It is encouraging that the light and velocity dispersion profiles
found in our simulated halos are in good agreement with available
observations of dSph and can also give some insight to understand new
puzzling observations.
}

Clearly, it is a long way to go from $z=8.3$ to the present day.
Therefore, in the rest of this paper, we are assuming that the stellar
component of our galaxies does not change after $z=8.3$ except due to
passive stellar evolution. This hypothesis, obviously, does not apply
to {\it all\/} observed dwarfs, but it may apply to {\it some\/} of
them - as we discuss later. The probability, $P$, of a galaxy
collapsed at redshift $z_{\rm coll}$ to survive to redshift $z$
without being incorporated into a larger object is approximately $P
\sim (1+z)/(1+z_{\rm coll})$ (Sasaki 1994). We can therefore estimate
that about 10\% of the galaxies present in our simulation at $z=8.3$
may have survived almost unspoiled until today. With this approach, we
cannot determine what dark matter mass these galaxies would have by
$z=0$: they can continue accreting\footnote{Note that a dark matter
  halo that virializes at $z=10$ by the time it reaches redshift $z=0$
  is about ten times larger in size and in mass even if it did evolve
  passively without accreting any satellite (\ie, in isolation). This
  is a consequence of the cosmological evolution of the Hubble
  constant and mean density of the IGM in which it is embedded (see Gunn
  \& Gott\ 1972; Ricotti \& Wilkinson\ 2004).}  the dark matter and
then lose a significant fraction of it during their orbital evolution
within the parent galaxy halo (in this case - Milky Way or Andromeda).
It is therefore important to keep in mind throughout this paper that
{\it we only use the properties of the stellar components of
  simulated galaxies\/}. Thus, we cannot make a direct test of the
tidal scenario, and have to rely on stars to tell us the story of the
dark matter.

In order to predict the luminosities and stellar densities of
simulated galaxies at $z=0$, we need to take into account stellar
mass-to-light ratios and stellar evolution. Since the simulated stars
would be old by $z=0$, we adopt a value for the mass-to-light ratio
$Y$ as measured in old galactic globular clusters, $Y\sim 1.5-2.0$
(e.g., Mandushev et al.\ 1991, Pryor \& Meylan 1993).  In addition,
after some $13\dim{Gyr}$ of stellar evolution, assuming Kroupa IMF we
estimated (using Starburst99, Leitherer et al.\ 1999) that our stars
will loose about 50\% of their mass. The exact value depends on the
assumed lower mass cutoff of the IMF.  Thus, in order to compute the
luminosities of the simulated galaxies at $z=0$, we adopt a fiducial
ratio of $M_*(z=8.3)/L_V(z=0)=3$. However, our results are not
immensely sensitive to that number - changing it to 2 or 10 would not
change our conclusions (see Fig.~\ref{figLR} and Fig.~\ref{figLRb}).

\section{Results}

\begin{figure}[t]
\plotone{\figname{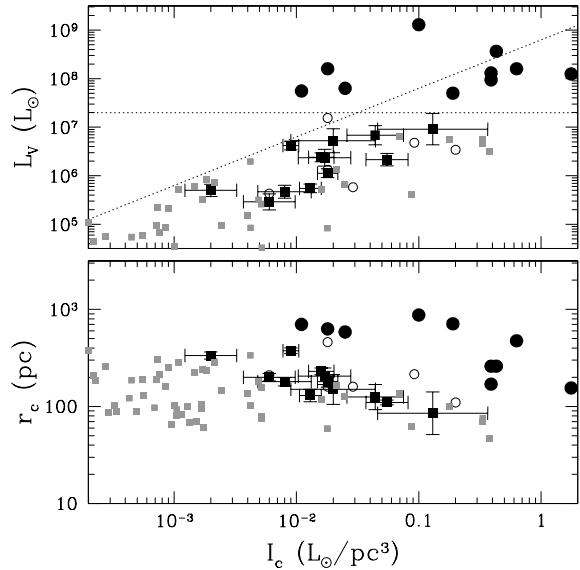}}
\caption{The V-band luminosity (top) and the core radius (bottom) vs
  the central luminosity density for the Local Group dwarfs (squares
  with error-bars, open circles, and filled circles - the different
  symbols are explained in the text) and for the simulated galaxies
  passively evolved to $z=0$ (gray squares). Two dotted lines show two
  possible boundaries between the ``fossils'' and \nameIIplu. The data
  is from Mateo (1998). We have complemented the data using updated
  distances and photometry from van den Berg (2000) and metallicities
  and metallicity spreads from Grebel et al. (2003).}
\label{figLR}
\end{figure}

The core of the reionization scenario is that some of the Local Group
dwarfs are true fossils of the pre-reionization era. Clearly, not all
dwarfs fall into this category, since many of them had recent episodes
of star formation. But is there a subset of Local Group dwarfs that
formed before reionization? We can attempt to answer this question by
comparing our simulated dwarfs as they would appear at $z=0$ with the
real observed galaxies. Figure~\ref{figLR} presents such a comparison
for the basic structural parameters: the luminosity $L_V$, the core
radius $r_c$, and the central luminosity density $I_c$, defined as
$I_c=\Sigma_V/2r_c$, where $\Sigma_V$ is the surface brightness. From
the top panel of that figure it is clear that the simulated galaxies,
shown by the gray squares without error-bars, do not exceed about
$10^7L_\odot$. Moreover, it appears that the observed galaxies (shown
by squares with error-bars, open circles, and filled circles - the
different symbols will be explained later in the text) also fall into
two groups - those with luminosities on the order of $10^7L_\odot$ and
below, and those with luminosities in excess of $5\times10^7L_\odot$.
In addition, these two groups are also clearly separated in $r_c-I_c$
plane.  The low luminosity dwarfs appear to trace the same
distribution as the simulated galaxies, having smaller core radii as
the central luminosity density increases. This observation can be
understood in terms of the less efficient cooling of the gas in dark
matter halos of smaller mass.
\begin{figure}[t]
\plotone{\figname{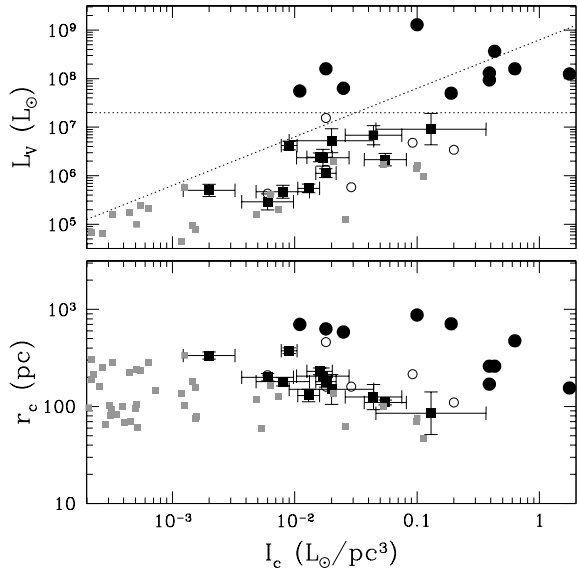}}
\caption{Same as in Fig.\
  \protect{\ref{figLR}} but assuming $M/L_V=10$ instead of the
  fiducial value $M/L_V=3$. The agreement between simulations and data
  is mildly sensitive to the assumes mass to light ratio for all the
  scaling relationships except the luminosity-metallicity relation
  that gets worst for larger $M/L_V$.}
\label{figLRb}
\end{figure}

Since our simulation has a finite size of the computational box, it is
clear that we miss most luminous galaxies forming at the same time as
our simulated dwarfs. The tilted dotted line roughly traces the upper
envelope of the simulated galaxies, so brighter galaxies that trace
the same distribution in the $I_c-L_V$ plane would fall below that
line. Some of the observed dwarfs are indeed in this region, but, as
the bottom panel of Figure~\ref{figLR} shows, their core radii are
somewhat larger than those of the simulated galaxies. Thus, we can be
confident that we do not miss a large fraction of more luminous dwarfs
in our simulation that follow the same distribution as the faint ones.
It does not, of course, mean that there should be no more luminous
galaxies in a simulation with a larger volume - they will be present,
but most of them will have masses in excess of $10^8\dim{M}_\odot$.
Therefore, because they will collapse via atomic line cooling, they
should have a very different internal structure and be much less
susceptible to losing their gas after reionization.
Figure~\ref{figLRb} is the same as Figure~\ref{figLR} but we assume
$M/L_V=10$ instead of the fiducial value $M/L_V=3$. The purpose of
this figure is to show that our results are robust and almost
independent on the assumed $M/L_V$ ratio.
 
\begin{figure}[t]
\plotone{\figname{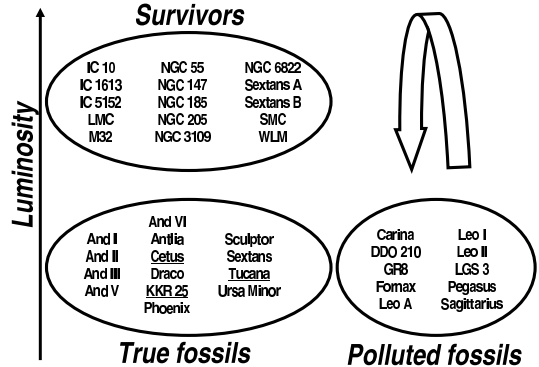}}
\caption{ A tentative separation of all Local Group dwarfs as well as
  other nearby dwarfs into three
  groups: \nameIIplu, that formed via the tidal scenario (mostly
  dIrr), ``true fossils'', that formed via the reionization scenario,
  and ``polluted fossils'', that went through both stages. In our
  tentative classification most dSphs (14) are true fossils: five are
  M31 satellites, six are Milky Way satellites and three (the
  underlined names) are isolated from both M31 and the Milky Way.}
\label{figFC}
\end{figure}

\hide{ In this framework these more luminous galaxies may have
  structural parameter that resemble dwarf ellipticals (dE) such as
  NGC205 and NGC147 or dIrrs. In this paper will group these two
  classes into the \nameIIplu, but should be noted that dEs have some
  structural properties consistent with being the high-luminosity end
  of the ``true fossils''.  }

It is tempting, therefore, to hypothesize that the two groups of low
and high luminosity dwarfs correspond to two distinct formation tracks
- the low luminosity ones are the ``fossils'' of the pre-reionization
era, and high luminosity ones formed their stars via atomic line
cooling perhaps via Kravtsov et al.\ (2004) scenario, as tidally
stripped more massive halos (in the absence of a better name, we will
call them \nameIIplu\ of reionization era).

\hide{ Note that the term \nameIIplu\ might be misleading since we do
  not expect that these galaxies might lose a very large fraction of
  their stars during the interaction with the Milky Way as this would
  be difficult to reconcile with the observed luminosity-metallicity
  relation. But it is plausible that gas compression can trigger star
  formation and tidal forces might heat the stellar component leading
  to changes of their structural parameters.  }

We can now see how our separation in these two groups (based only on
the luminosity) compare with the available data on the star formation
history of observed dwarfs. Hereafter we will adopt the star formation
histories compiled by Grebel \& Gallagher\ (2004). It is well known
that almost all dwarfs in addition to a more or less prominent old
population clearly show some more recent star formation. Therefore
taken at face value, our hypothesis does not hold in comparison with
the data. However, it is unreasonable to assume that {\it all\/} of
the galaxies that formed during the pre-reionization era formed
absolutely no stars afterwards. Some of them should surely continue
accreting dark matter, becoming more massive and switching from the
``reionization'' track to the ``tidal'' track. To be more specific,
the ability of a dark matter halo to accrete fresh intergalactic gas
is controlled by the filtering mass (Gnedin 2000b). In the standard
CDM$+\Lambda$ theory the filtering mass increases with time, as does
the mass of an individual dark matter halo.  Thus, the question of
whether a given halo is able to accrete fresh gas after reionization
(and, presumably, form new stars) depends on whether its mass grows
faster or slower than the filtering mass. We can, therefore, expect
that some of galaxies that formed before reionization will form no new
stars, and some of them will form more stars as they fall into the
Milky Way or Andromeda halo and their local filtering mass and
accretion history vary.
\new{The specific rate with which the gas is re-accreated must depend
  on the details of a satellite orbit, as has been emphasized by
  Kravtsov et al.\ (2004), and so a large variety of possible star
  formation histories is expected, in a qualitative agreement with
  the observational data. A much more advanced simulation than the
  one presented here will be required, however, to check whether a
  quantitative agreement between the theory and the data holds -  for
  example, whether the relative fractions of dwarfs with different
  star formation histories are consistent with the data.}

As mentioned in the introduction another problem faced by the
``fossil'' hypothesis is the large metallicity spreads observed in
dwarfs with a dominant old population. Grebel \& Gallagher\ (2004),
have suggested that these large spreads are not compatible with star
formation protracted over a period of less than 1 Gyr (the time elapsed
before reionization). They conclude therefore that star formation in
old dwarfs was not affected by reionization. In defense of the
``fossil'' hypothesis we will show as part of our analysis that the
metallicity spreads in our simulated dwarfs are consistent with
observations. Note that star formation in these dwarfs was protracted
for less than 1 Gyr.  Our result is probably due to accretion of stars
from satellites with different metallicities and wouldn't be captured
by any standard chemical evolutionary model.

In order to schematically separate the observed dwarfs in groups
according to their star formation histories, we call those dwarfs that
formed few (say, less than 30\%) stars after reionization ``true
fossils'', and those that did form a substantial amount of stars
``polluted fossils''.  The various symbols in Figure~\ref{figLR} and
the following ones correspond to these classes: ``true fossils'' are
shown with filled squares with error bars, ``polluted fossils'' are
shown with open circles. Galaxies with dominant younger stellar
populations (they mostly are dIrrs) are shown with larger filled
circles and we call them \nameIIplu.  For the last two groups the
error bars are not shown for the sake of clarity.  All the galaxies in
our simulation have almost identical properties to the ``true
fossils'' and most ``polluted fossils''. All these galaxies have a
prominent old stellar population as expected. Instead all the galaxies
with a prominent young population (filled circles) are part of a
distinct population that is not present in our simulation.

Figure \ref{figFC} presents a separation of the Local Group
dwarfs into these three types. Galaxies identified as \nameIIplu\ can
be thought of as those dwarfs which formed after reionization (or
mostly after reionization), and for which the total mass exceeded the
local filtering mass at some moment during their evolution. Polluted
fossils formed a significant fraction of their stars in the
pre-reionization era, but also exceeded the local filtering mass at
some point in their history.  The ``true fossils'' formed most of
their stars in the pre-reionization era. As a weak argument in favor
of this hypothesis, we notice that ``polluted fossils'' are on average
brighter than the ``true'' ones.
\begin{figure}[t]
\plotone{\figname{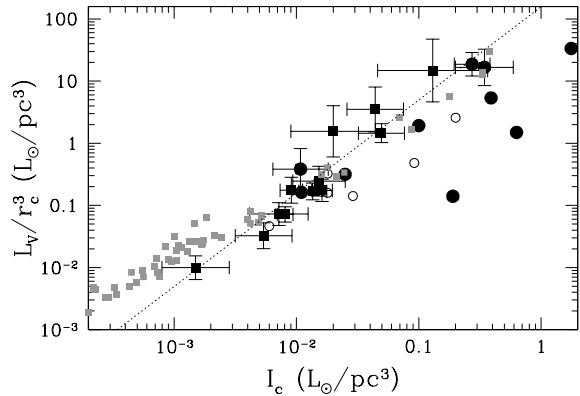}}
\caption{The V-band luminosity per cube of the core radius vs
the central luminosity density for the Local Group dwarfs 
and for the simulated galaxies (gray squares). In this and further figures
filled black squares mark ``true fossils'', open circles mark ``polluted
fossils'', and filled circles show \nameIIplu. The dotted line
shows the power-law fit to the ``true fossils'' with 
the slope fixed to $3/2$ (see Gnedin 2000a for details).}
\label{figII}
\end{figure}

This separation of galaxies into individual classes is in some cases
rather artificial. For example both Sextans and Fornax had some recent
star formation. We labeled Sextans a ``true fossil'' because this
recent episode was very small, while Fornax had a factor of 3 to 4
more stars formed after reionization, and so is labeled a ``polluted
fossil''. Based on their star formation history Leo A and Leo I could
also be classified as low luminosity \nameIIplu.




In order to support our conclusion, we consider other properties of
the simulated and real galaxies.  Figure \ref{figII} shows the
comparison between the data and the simulation for the relation
between $L_v/r_c^3$ and $I_c$, discussed in Gnedin (2000a). We find a
scaling relationship 
$$
\frac{L_v}{I_cr_c^3} \approx 
100 \left(\frac{I_c}{1 L_\odot\dim{pc}^{-3}}\right)^{0.5},
$$
shown by the dotted line. If the interpretation of
Gnedin (2000a) of this relation is correct, then it simply reflects
the Schmidt law for star formation on scales of hundreds of parsecs -
in that case the agreement between the data and the simulation is by
construction, rather than a success of the model: the star formation
in the simulation is assumed to follow the Schmidt law (1.5 slope in
the relation between the star formation rate and gas density), and the
amplitude is fixed by observational constraints. On the other hand,
this agreement does support the reionization scenario, because 
\nameIIplu\ do not follow this tight relation, which could be expected
if their stars formed in several independent bursts.
\begin{figure}[t]
\plotone{\figname{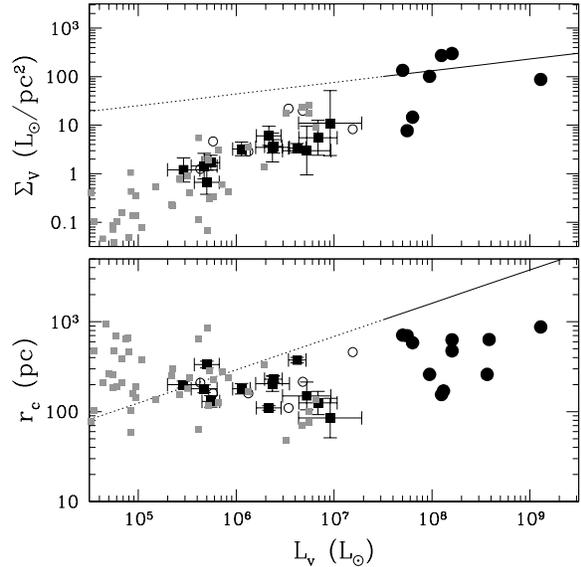}}
\caption{Surface brightness (top) and core radius
  (bottom) as a function of the V-band luminosity for the simulated
  and observed dwarfs. Symbol markings are the same as in Fig.\ 
  \protect{\ref{figII}}. For comparisons we show (solid lines) the
  scaling relationships for the more luminous Sc-Im galaxies ($10^8L_\odot
  \simlt L_B \simlt 10^{11}L_\odot$) derived by Kormendy \& Freeman\ 2004.}
\label{figL}
\end{figure}

Note that $L_v/(I_c r_c^3)$ is the ratio of the total luminosity to
the core luminosity and is therefore a measure of how concentrated 
the light of the dwarf galaxy is. This ratio is less than 10 for dSph
galaxies like Draco and Ursa Minor ($I_c \sim 5 \times 10^{-3}L_\odot
\dim{pc}^{-3}$) and is almost 100 for dEs like NGC205 and NGC147 
($I_c \sim 1L_\odot\dim{pc}^{-3}$). The galaxies that we classified as
\nameIIplu\ have generally central luminosity density comparable to dEs 
but less concentrated light profiles. We do not observe such objects
in our simulation - despite the fact that cores of these galaxies, if they
were present in the simulation, would be completely resolved.

Figure \ref{figL} compares the surface brightness and the core radii
of simulated and observed galaxies as a function of the luminosity.
The simulation reproduces very well the structural properties and
scaling relations of the galaxies that we classified as ``true
fossils''. The lines show, for comparison, the scaling relations
derived by Kormendy \& Freeman\ (2004) for more luminous late-type
galaxies (Sc-Im) with luminosities in the range $10^8L_\odot \simlt
L_B \simlt 10^{11}L_\odot$.

\begin{figure}[t]
\plotone{\figname{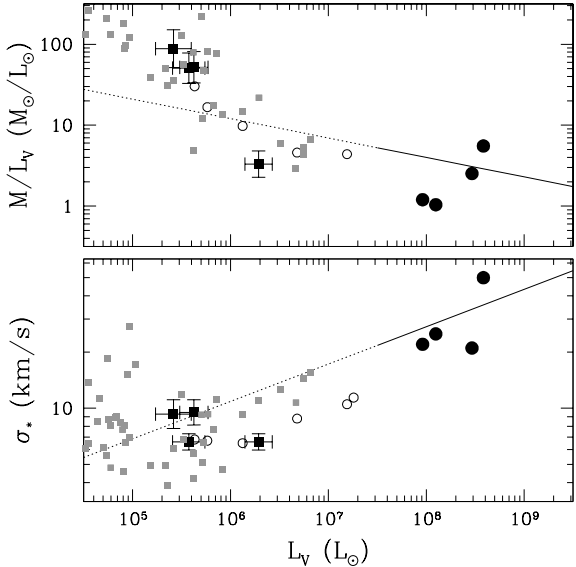}}
\caption{Total mass-to-light ratio (top) and stellar velocity dispersion
  (bottom) for the simulated and observed dwarfs. Symbol markings and
  the meaning of the lines are the same as in Fig.\ \protect{\ref{figII}}. Fewer data
  points appear in this figure because the data do not exist for all
  Local Group dwarfs.}
\label{figSL}
\end{figure}
Figure \ref{figSL} compares the dynamical state of these galaxies: the
stellar velocity dispersion and the total mass-to-light ratio. The
latter quantity for the simulated galaxies is computed in the same way
as for the observed ones ($M/L = 136\sigma_*^2r_c/L$, see Mateo 1998),
and is not taken from the simulations, because, as we have mentioned
above, there is no simple way to relate the dark matter mass of
simulated galaxies at $z=8.3$ with their dark matter masses at $z=0$.
For a comparison, the solid lines show the same scaling relations for
late-type galaxies (Kormendy \& Freeman\ 2004).  Within the errors the
$\sigma-L_v$ relation seems to be the same power law for all the
galaxy types, but ``true fossils'' obey a steeper relationship between
the mass-to-light ratio and luminosity than dIrr and Sc-Im galaxies.
\hide{Note that the large $M/L$ ratio in our simulated galaxies is not
produced by the feedback from SN explosions as it is usually assumed.
In our simulations radiative feedback from ultraviolet radiation is
sufficient to reproduce observations. In particular, the feedback is
the combination of FUV radiation that inhibit an efficient star
formation by dissociating \H2 (the main coolant) and EUV radiation
that ionizes and heats the gas producing mass loss trough strong
galactic winds.}

\begin{figure}[t]
\plotone{\figname{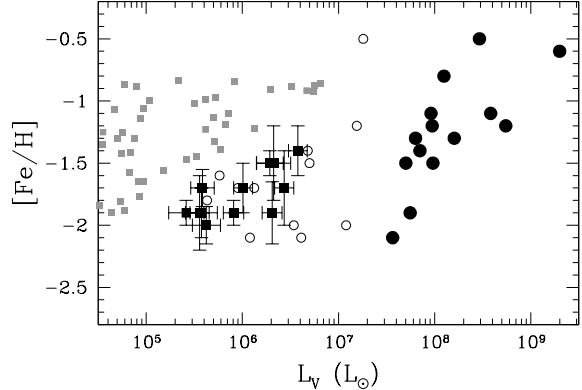}}
\caption{The stellar metallicity for the simulated and
  observed galaxies. Symbol markings are the same as in Fig.\
  \protect{\ref{figII}}. We adopt solar abundances and yield
  y=$\rho_Z/\rho_*=1$ Z$_\odot$.}
\label{figZL}
\end{figure}
Figure \ref{figZL} gives yet another comparison between the simulation
and the data: the metallicity as a function of stellar luminosity. 
The agreement between stellar metallicities in the simulation and the
data is somewhat worse: while the spread in the distribution is
consistent with the data, simulated galaxies are more metal-rich by
about a factor of $1.5-2$. At this point we are not worried by this
disagreement as our simulation is not sophisticated enough to predict
the stellar metallicities to within a factor of two.  There are several
physical and numerical effects that will lead to overestimated stellar
metallicities in the simulations, such as preferential ejection by
supernovae of metal-enriched gas (MacLow \& Ferrara 1999), imperfect
mixing of the supernova ejecta on scales not resolved by a simulation,
reduced gas accretion on the simulated galaxies due to a limited size
of the computational box, etc.

Low metallicity stars can be produced in two different ways: either by
inefficient star formation combined with continuous gas loss, or in a
single short burst.  The first scenario is consistent with the
properties of dSph galaxies which have an extended low surface
brightness stellar component. The second scenario is usually
associated with a higher efficiency of star formation, which results
in a more concentrated light profile than the one observed in dSph,
and is probably more appropriate to explain the metal enrichment of
globular cluster and dwarf Elliptical galaxies.

\begin{figure}[t]
\epsscale{1.6}
\plottwo{\figname{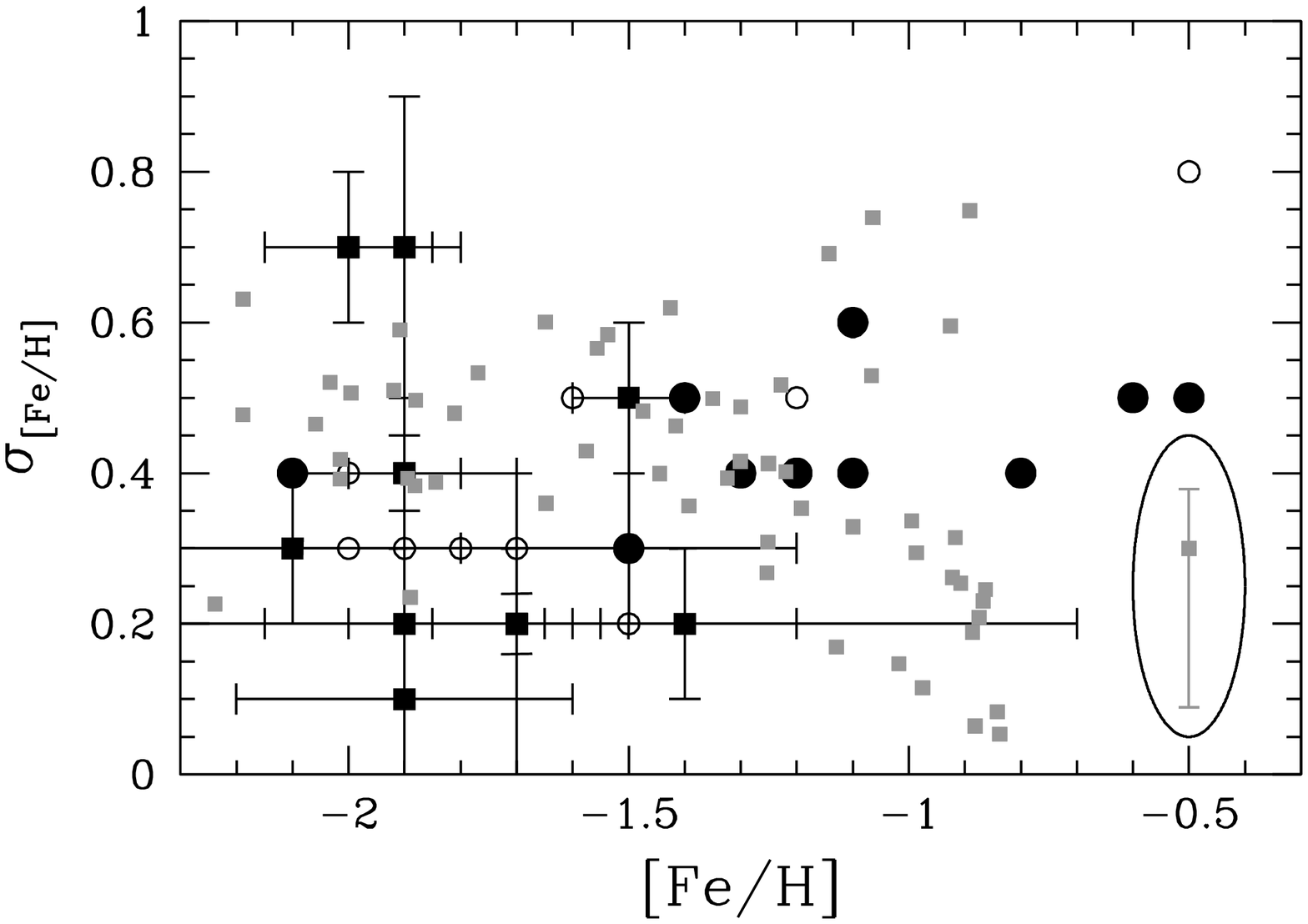}}{\figname{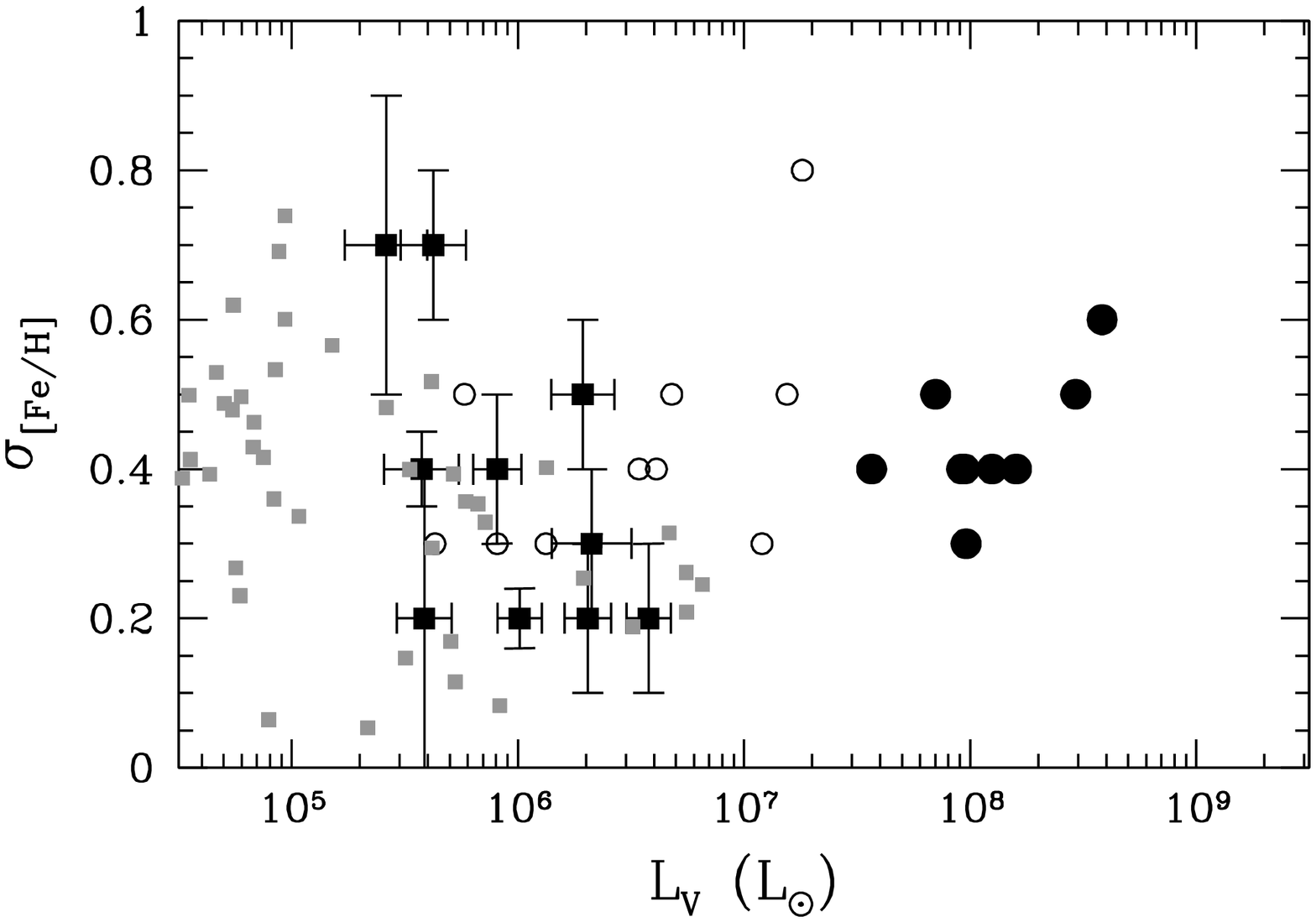}}
\caption{Metallicity spread of the stellar populations
  $\sigma_{[Fe/H]}$, as a function of mean metallicity [Fe/H] for the
  simulated (gray squares) and observed dSphs (symbol markings are the
  same as in Fig.\ \protect{\ref{figII}}). The spread
  $\sigma_{[Fe/H]}$ is defined as the variance 
  of the number weighted metallicity distribution of the ``stellar
  particles'' in each dwarf. A rough estimate of the uncertainty on
  the metallicity spread is shown by the circled errorbar on the top
  left corner. The error is estimated as the mean difference between
  $\sigma_{[Fe/H]}$ using the mass weighed and the number weighed
  metallicity distributions.}
\label{figZs}
\end{figure}
We have already anticipated that our simulation reproduce the large
metallicity spreads observed in dSphs. In Figure~\ref{figZs} we show a
comparison between the simulated and the observational data for the
metallicity spreads, $\sigma_{[Fe/H]}$, both as a function of the mean
metallicity [Fe/H] (left panel) and as a function of luminosity $L_V$
(right panel). We find that the typical metallicity variance in each
dwarf is about 0.5 dex, in good agreement with observations. The star
formation history in simulated dwarfs is characterized by multiple
instantaneous starbursts extended over a period of 0.5-1 Gyr. The
starbursts often do occur almost the same time but in distinct dark
matter sub-halos that are in the process of merging to form a more
massive dwarf. The accretion of stars from the hierarchical
distribution of sub-halos partially explains the large metallicity
spread that we find in our simulation.

Thus we conclude that the large metallicity spread observed in the old
stellar population of dSphs is not a definite signature of star
formation extended over 2-4 Gyrs (Ikuta \& Arimoto 2002; Grebel \& Gallagher
2004).  Neither the argument based on the long timescale (\eg, 1-4
Gyr) needed to have a substantial iron production by Type Ia SN is a
strong argument in favor of prolonged star formation. The typical time
scale for iron enrichment by Type Ia SN is uncertain and very likely
is dependent on the mode of star formation. Matteucci \& Recchi (2001)
estimated that this time scale varies from 40-50 Myr for an
instantaneous starburst to 0.3 Gyr for a typical elliptical galaxy to
4-5 Gyr for a disk of a spiral galaxy like the Milky Way. The often
quoted 1 Gyr time scale is based on observations of [$\alpha$/Fe]
versus [Fe/H] in the solar neighborhood stars and is not an universal
value. A more detailed study would be needed to derive accurate metal
abundance ratios based on the star formation and merger histories
found in the simulations. This is beyond the aim of the present paper
but it is worth pursuing in a separate study.

\hide{ It is important to notice that dwarf galaxies in which a
  substantial fraction of the stars are lost due to tidal effects
  would decrease their luminosity but not the metallicity of their
  stars producing an observable shift in the luminosity-metallicity
  relation of the same magnitude for both Fe and O. Since from figure
  \ref{figZL} appears that the offset of the $L_V-Z$ relation is
  different for O and Fe, this scenario does not look promising to
  explain the metallicity of dSph galaxies.  }
\begin{figure}[t]
\epsscale{1.0}
\plotone{\figname{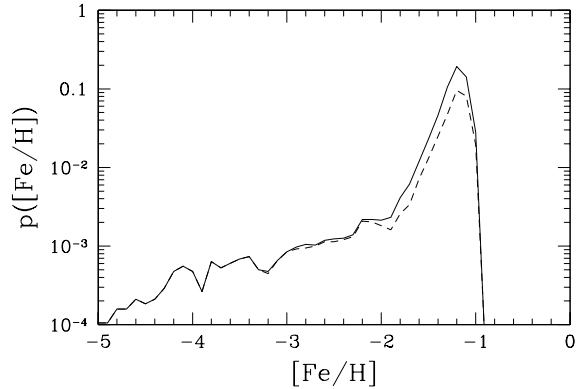}}
\caption{The stellar metallicity distribution of one simulated
  galaxy. The solid and dashed lines show the number and mass weighted
  distributions respectively.}
\label{figZhi}
\end{figure}
We can also study the stellar metallicity distribution in each one of
our simulated dwarfs. Figure~\ref{figZhi} show such an example. The
solid and dashed lines show the number and mass weighted distributions
respectively. The metallicity distribution is skewed toward low
metallicities but only a small fraction of the stars, having
metallicities $<10^{-4}$ Z$_\odot$, could be defined as Population~III
stars.

\section{Discussion}

In this paper we have proposed that dwarf galaxies of the
Local Group (and, by implication, all other dwarf galaxies in the universe)
formed in three different evolutionary paths:
\begin{description}
\item[``true fossils''] formed most of their stars in the pre-reionization
  era, and have had little (say, less than 30\%) star formation since then;
\item[``polluted fossils''] started as true fossils, but had substantial
  episode(s) of subsequent star formation as they continued accreting mass
  and were tidally shocked during the formation of the parent galaxy halo;
  finally
\item[\nameIIplu] started forming stars mostly after reionization.
\end{description}

This conclusion is motivated by considering star formation histories
of Local Group dwarfs and by comparing their properties with
properties of simulated galaxies that formed all their stars before
cosmological reionization. The fact that our simulated galaxies are
remarkably similar to the subset of the Local Groups dwarfs that we
identified as ``true fossils'' in Figure~\ref{figFC} renders support
to our conclusions. This subset includes about all known dSphs.

To avoid confusion it is worth noting that the structural properties
of the galaxies that we call ``true fossils'' are the result of
feedback processes in action prior to reionization.  The effect of
reionization is simply to preserve those properties by suppressing the
star formation rate in halos with masses that remain smaller than the
filtering mass in the IGM. In this sense our model differs from the
standard ``reionization'' scenario for the formation of dwarf
galaxies.

How would one test this hypothesis? For example, the main argument of
Kravtsov et al.\ (2004) for the tidal origin of the majority of
Local Group dwarfs is
their preferential location close to the parent galaxy. However, true
fossils, living in the oldest dark matter halos of their mass, are
highly biased, and are also preferentially located closer to the
center of the parent halo. Moreover there are at least two examples of
dSph galaxies that are quite far from massive galaxies:  Cetus and
Tucana, and perhaps the newly discovered Apples1 (Pasquali et al.\ 2005).

A more powerful test is based on our conclusion that \nameIIplu\ form
later than the ``true fossils''. While measuring absolute ages of
dwarf galaxies to $1\dim{Gyr}$ precision is not possible at the
moment, determining the {\it relative\/} difference of about
$1\dim{Gyr}$ in age between two galaxies observed by a uniform
technique may, in fact, be possible for a specific range of
metallicities and ages using the so-called ``horizontal branch index''
(G.\ Gilmore, private communication; also see Zinn 1993; Harbeck et
al. 2001; Makey \& Gilmore 2004): younger galaxies have redder
horizontal branch for the same metallicity,
\new{although, how precise the horizontal branch index method is and
  the role of the so-called ``second parameter effect'' is
  still a matter of debate (the anonymous referee, private communication).}

Luckily, some of the Local Group dwarfs fall into that range of
parameters. For example, two galaxies of our ``true fossil'' group -
And I and And II - have $[{\rm Fe}/{\rm H}]\approx-1.5$, similar to
the iron abundance in two of \nameIIplu\ - WLM and NGC3109 (having
pairs of galaxies with the same metallicity is absolutely crucial as
even small differences in metallicities can obscure large differences
in age). Thus, we can make a prediction that most of the stars that
formed in the first burst of star formation in WLM and NGC3109 should
be about $1.5\dim{Gyr}$ younger than most of stars in And I and And
II. The color-magnitude diagrams of And I, And II, and WLM are
measured down to $V>26$, well below the horizontal branch (Da Costa,
Armandroff, \& Caldwell 2002; Rejkuba et al.\ 2000), although, to the
best of our knowledge, no photometry down to that levels exists for
NGC3109.  Visual comparison of the horizontal branch morphologies for
the first three galaxies indicate that And I and And II are indeed
substantially older than WLM, also a more rigorous, quantitative test
is required to confirm the visual impression. Such a comparison would
be a crucial test of our hypothesis - if it is found that And I and
And II are not substantially older than WLM and NGC3109, that would
imply that they are not ``true fossils'', and by inference, few other
dwarf spheroidals are. The main problem that affects this method when
comparing the old populations of dSphs to dIrrs is that normally dIrrs
have mixed populations and the overlap between the blue horizontal
branch and the young main sequence make it difficult to derive
reliable horizontal branch morphology indices in dIrrs.  Perhaps a
more reliable and accurate relative age indicator is the main
sequence turn-off. The internal accuracy can be better than 1 Gyr
in few nearby dwarfs with photometric data that extends below the
main-sequence turnoff.

Using available data, Grebel \& Gallagher (2004) address the impact of
reionization on the star formation histories of dwarf galaxies. They
conclude that most or all dwarf galaxies present an old population
that is coeval with the oldest Milky Way populations. They also find
that all of the Local Group dwarf galaxies exhibit considerable
abundance spreads even within their old populations indicating that
star formation did not occur in a single burst, but in episodes that
must have extended over several Gyr, well beyond reionization. In this
paper we have argued that the observed metallicity spreads are not
necessarily an indication of star formation extended over several
Gyr. In our simulations all the dwarfs that form before reionization
have metallicity spreads consistent with observations. The large iron
abundances in dSphs have also been used as an argument in favor of an
extended star formation. This is because the time-scale for iron
enrichment is often assumed to be longer than 1 Gyr. But we point
out that this assumption may be true for a continuous mode of star
formation, but is not necessarily 
universally true. The time scale depends on the
star formation history and could be as short as $50\dim{Myr}$ for a
bursting mode of star formation (Matteucci \& Recchi 2001).

In principle, $[\alpha/{\rm Fe}]$ ratio could be used to discriminate
between star formation histories consisting of a single burst or a
more prolonged episode of star formation. At the moment it is not
clear how much this test could distinguish ``true fossils'' from
\nameIIplu, since both are expected to have the main burst of star
formation for a relatively short period of time. The star formation
rate in the ``true fossils'' should decrease substantially after
reionization. In addition our simulations show that they have a very
peculiar mode of star formation characterized by multiple short bursts
(about $10-100\dim{Myr}$ long).  It is unclear to us how this mode of
star formation may affect their $[\alpha/{\rm Fe}]$ ratio and be used
as a discriminant. Using existing data on $[\alpha/{\rm Fe}]$ ratios
in dSphs (Tolstoy et al.\ 2003, Venn et al. 2004) it might be possible
to distinguish between different star formation and enrichment
histories but this would require a more detailed modelling that is
beyond the aim of the present paper.

In addition, we find that the most massive end of the ``true fossils''
population in our simulation (with mass $M>1-2\times10^8\dim{M}_\odot$)
retain a fraction of their gas even after reionization (see Fig.~1). These few
``transition'' objects may continue forming stars for a while after
reionization and have a more extended period of star formation with
respect to the rest of the ``fossil'' population.


\begin{figure}[t]
\epsscale{1.0}
\plotone{\figname{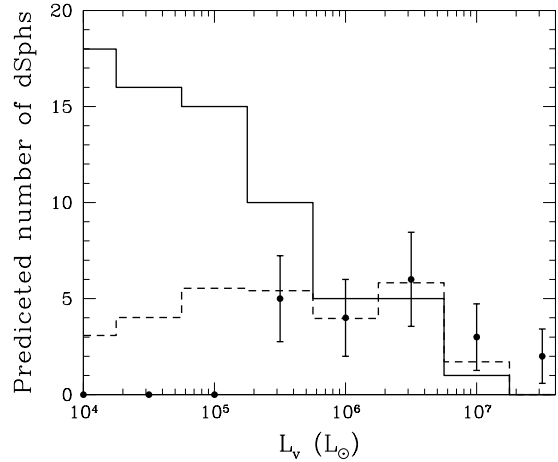}}
\caption{Prediction of the number of faint dSphs per logarithmic luminosity
bin in the Local Group (histograms). The points with error bars (Poisson 
errors) show the number of observed dwarf galaxies in the Local Group.
The solid line refers to the case in which the survival probability of
reionization fossils is constant as a function of luminosity, while
the dashed line shows an example of a case in which the survival
probability is mildly luminosity dependent (see text for details).
}
\label{figpre}
\end{figure}

To conclude, we would like to emphasize another exciting possibility. From
the results of our simulations, we feel quite confident in predicting
the existence of fainter and lower surface brightness galaxies than
the one currently discovered (see the gray squares at the extreme left
corners in Figs.~3, 6 and 7). Two promising candidates for such a class
of dwarf galaxies were recently discovered in M31 and near the Milky Way
(Zucker et al.\ 2004, Willman et al.\ 2005).  

As we mentioned above, it is
mostly likely that the majority of reionization fossils will merge into
larger objects by $z=0$, and only a small fraction of them will
survive. Thus, we can introduce a ``survival probability'' $S$ (which
can be a function of other parameters, in particular the luminosity)
of a reionization fossil to remain intact by the present time. This
probability can be estimated by comparing the number of observed
dwarfs in the Local Group (the total mass is about
$4\times10^{12}\dim{M}_\odot$) with the number of simulated galaxies
in our volume (with the total mass of
$1.2\times10^{11}\dim{M}_\odot$). If we assume that both the
observations and our simulation are complete for galaxies with $L_V$
between $10^6\dim{L}_\odot$ and $3\times10^6\dim{L}_\odot$, we find a
value of $S=3\%$ for the survival probability in this luminosity range.

Figure \ref{figpre} shows our predictions for the abundance of these
``ultra-dwarfs'' in the Local Group. The solid line in Fig.\
\ref{figpre} refers to the case in which the 
survival probability of reionization fossils is kept constant at $3\%$
- in this 
case the observations may be missing about half of the satellites with
luminosities in the range $2\times10^5\dim{L}_\odot \simlt L_V \simlt
6\times10^5\dim{L}_\odot$ (due to a small size of the computational
box, out simulation is incomplete for $L_V \simgt 10^7\dim{L}_\odot$,
as we discussed above). Alternatively, we can assume that the
observations are complete in this luminosity range, in which case the
survival probability must be a weak function of luminosity. Fitting the
power-law dependence (as an example) to the abundance of observed
dwarfs in the three lowest luminosity bins ($2\times10^5\dim{L}_\odot
\simlt L_V \simlt 6\times10^6\dim{L}_\odot$), we find $S(L_V)=3
\%\times(L_V/2\times 10^6\dim{L}_\odot)^{1/3}$ (this fit reaches
$S=100\%$ for the Milky Way luminosity, and so, at least, makes
sense). This case is shown with the 
dashed line in Fig.\ \ref{figpre}. As one can see, a modest change in
the survival probability makes a large change in the predicted numbers
of dwarf galaxies in the $L_V\simlt10^5\dim{L}_\odot$ luminosity
range, and, thus, predicting the precise abundance of ``ultra-dwarfs''
will be difficult. On the other hand, measuring the abundance of dwarf
galaxies in 
this luminosity range will put non-trivial constraints on the
luminosity dependence of the survival probability, and, therefore, on
the hierarchical merging process and galaxy clustering on extremely
small mass scales. 

Although the detection of such objects in the Milky Way is
very difficult if not impossible using photometric techniques (due to
contamination from foreground stars), in the future it might become
possible to search for these objects using spectroscopic techniques by
measuring the radial velocities of many stars. Any observation able to
put constraints on properties of the faintest bound object in the
Local Group would be of extreme interests not only for understanding
galaxy formation in the early universe, but, more importantly, for
uncovering the properties of the elusive dark matter particles.

\acknowledgements We are grateful to the anonymous referee for very
useful comments that led to great improvement of the original
manuscript. We thank R.\ Wyse, G.\ Gilmore, A.\ Kravtsov, O.\ Gnedin
and J.\ P.\ Ostriker for valuable discussions and important insights.
MR was supported by a PPARC theory grant. This work was supported in
part by HST Theory grant HST-AR-09516.01-A, NSF grant AST-0134373, and
by National Computational Science Alliance under grant AST-020018N and
utilized IBM P690 array at the National Center for Supercomputing
Applications.

\end{document}